\documentclass[12pt]{article}

\usepackage{newtxtext,newtxmath}

\usepackage{graphicx}

\usepackage[letterpaper,margin=1in]{geometry}

\renewenvironment{abstract}
	{\quotation}
	{\endquotation}

\date{}

\makeatletter
\renewcommand{\fnum@figure}{\textbf{Figure \thefigure}}
\renewcommand{\fnum@table}{\textbf{Table \thetable}}
\makeatother

\usepackage{scicite}

\usepackage{url}

\def\scititle{
	Thermally quenched metastability in metal–insulator transitions via elemental substitution
}
\title{\bfseries \boldmath \scititle}

\author{
	Hiroshi~Oike$^{1\ast}$,
	Yasunori~Takahashi$^{2}$,
	Keisuke~Shibuya$^{3}$,
	Masaki~Nakano$^{4}$,\and
	Tatsuki~Hanada$^{2}$,
	Motoaki~Hirayama$^{5}$,
	Hiroko~Tokoro$^{6}$,
	Fumitaka~Kagawa$^{5,7}$\and
	\small$^{1}$Research Center for Materials Nanoarchitectonics,
	National Institute for Materials Science, Tsukuba, Japan.\and\
	\small$^{2}$Department of Applied Physics,
	The University of Tokyo, Bunkyo-ku, Japan.\and\
	\small$^{3}$National Institute of Advanced Industrial Science and Technology,
	Tsukuba, Japan.\and\
	\small$^{4}$Graduate School of Engineering and Science,
	Shibaura Institute of Technology, Koto-ku, Japan.\and\	
	\small$^{5}$RIKEN Center for Emergent Matter Science,
	Wako, Japan.\and\
	\small$^{6}$Department of Materials Science,
	University of Tsukuba, Tsukuba, Japan.\and\	
	\small$^{7}$Department of Physics,
	Institute of Science Tokyo, Meguro-ku, Japan.\and	
	\small$^\ast$Corresponding author. Email: OIKE.Hiroshi@nims.go.jp
}
\begin{document} 

\maketitle

\begin{abstract} \bfseries \boldmath
Thermal quenching inhibits equilibration toward the thermodynamic ground state during phase transitions, revealing metastable phases such as structural glasses and quenched alloys. Whether such thermally quenched metastability can be realized in metal–insulator transitions has remained an open question because these transformations are governed by collective electronic reorganization rather than atomic diffusion. We demonstrate that rapid cooling exceeding 10$^9$ K s$^{-1}$ kinetically avoids the metal–insulator transition, stabilizing a long-lived metastable metallic phase in tungsten-substituted VO$_2$. Temperature-dependent relaxation reveals nucleation-dominated kinetics with a thermal activation barrier introduced by tungsten substitution. Our results establish elemental substitution as a route to thermally quenched metastability in metal–insulator transitions, expanding metastable phase control to electronic phases.
\end{abstract}

\noindent
In liquids or solid solutions, slow cooling often results in crystallization or phase separation at low temperatures. Under rapid cooling, however, high-temperature phases can be supercooled because phase transitions generally require a finite time to complete. When such supercooled states are maintained down to sufficiently low temperatures, atomic motion effectively freezes on experimental timescales (e.g., seconds), and the system remains trapped in a metastable state. Such thermally quenched metastability has long been exploited to develop materials such as structural glasses and hardened steels. In these systems, phase transitions rely on atomic or molecular diffusion, and the suppression of such motion is directly linked to the emergence of metastability \cite{angell1995formation, debenedetti2001supercooled}.

Recently, thermal quenching has led to the discovery of metastable electronic phases in a wide range of correlated materials \cite{kagawa2013charge, oike2015phase, sato2017electronic, sasaki2017crystallization, yoshida2014controlling, oike2016interplay, karube2016robust, berruto2018laser, birch2019increased, matsuura2021kinetic, matsuura2023thermodynamic, katsufuji2020nucleation,oike2018kinetic}. For instance, in the organic conductors $\theta$-(BEDT-TTF)$_2$X and the layered dichalcogenide 1T-TaS$_2$, which exhibit metal--insulator transitions accompanied by charge ordering under slow cooling \cite{mori1998systematic, wilson1975charge}, rapid cooling freezes nonequilibrium charge configurations, resulting in metastable metallic states with resistivities orders of magnitude lower than those of the charge-ordered phases \cite{kagawa2013charge, oike2015phase, sato2017electronic, sasaki2017crystallization, yoshida2014controlling}. Thermally quenched metastability further applies to spin and orbital degrees of freedom, as exemplified by magnetic skyrmions \cite{oike2016interplay, karube2016robust, berruto2018laser, birch2019increased}, ferromagnetic phases \cite{matsuura2021kinetic, matsuura2023thermodynamic}, and an orbital-disordered phase \cite{katsufuji2020nucleation}, and leads to superconductivity in IrTe$_2$, whose equilibrium ground state is non-superconducting \cite{oike2018kinetic}. In parallel with the exploration of metastable electronic states, extensive effort has been devoted to understanding the mechanisms of metastability in electronic systems, where long-range diffusion of atoms or molecules is absent. In triangular-lattice systems such as $\theta$-(BEDT-TTF)$_2$X and 1T-TaS$_2$, metastable metallic phases have been discussed in connection with competing charge-ordering patterns arising from geometrical frustration \cite{kagawa2013charge, stojchevska2014ultrafast, vaskivskyi2015controlling}. However, whether geometrical frustration is essential for the metastability of a metallic phase remains unclear.

In the present study, we focus on tungsten-doped vanadium dioxide, V$_{1-x}$W$_x$O$_2$, for three reasons. First, VO$_2$ is not a triangular-lattice system; its metal–insulator transition involves a rutile-to-monoclinic structural change accompanied by V–V dimerization. Consequently, the lattice-geometry-driven competition seen in $\theta$-(BEDT-TTF)$_2$X and 1T-TaS$_2$ is absent. Second, the electronic phase diagram of V$_{1-x}$W$_x$O$_2$ (Fig.~\ref{fig:quenching}A) \cite{shibuya2010metal} suggests energetic competition between two phases, particularly in the composition range $x = 0.05$--$0.15$. In Ising models with a similar phase-diagram structure \cite{oike2025thermally}, ordering kinetics becomes slow when competing phases are nearly degenerate, implying that targeting such phase-competition regions may enable access to metastable states within experimentally achievable cooling rates. Third, recent experiments point to the presence of a metallic state hidden behind the insulating ground state, although they are distinct from thermal quenching experiments. For example, pulsed-laser excitation has been shown to induce transient metallic behavior in VO$_2$ that persists for several hundred picoseconds \cite{hilton2007enhanced, morrison2014photoinduced}, and continuous X-ray exposure gradually drives V$_{1-x}$W$_x$O$_2$ into a metallic state \cite{shibuya2011x}. Thus, V$_{1-x}$W$_x$O$_2$ is well suited for investigating thermal-quench-induced metastable phases in the absence of geometrical frustration.

\subsection*{Emergence of a thermally quenched metastable metallic state}

\begin{figure} 
	\centering
	\includegraphics[width=0.6\textwidth]{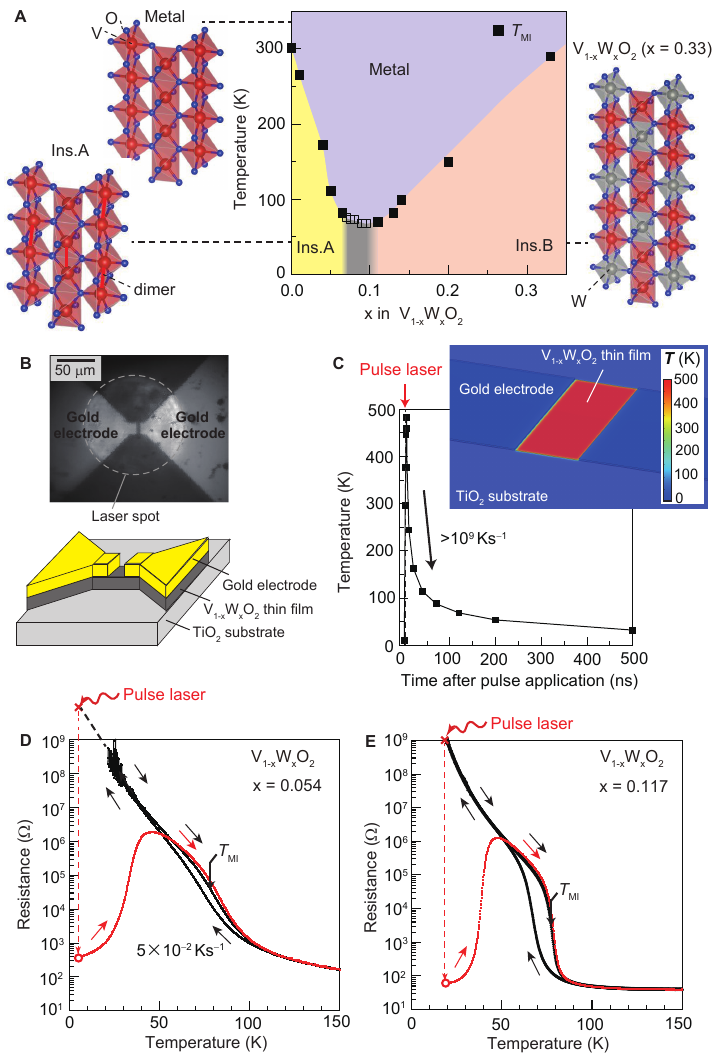} 

	\caption{\textbf{Thermally quenched metastable metallic phase in V$_{1-x}$W$_x$O$_2$.}
		(\textbf{A}) Temperature–$x$ phase diagram of V$_{1-x}$W$_x$O$_2$, constructed using reported data \cite{shibuya2010metal}. Ins. A denotes an insulating phase accompanied by V–V dimerization, whereas Ins. B denotes an insulating phase without V–V dimerization \cite{okuyama2015x}. Crystal structures were visualized using VESTA \cite{momma2011vesta}, based on reported crystallographic data \cite{longo1970vo2,israelsson1970phase}. (\textbf{B}) Optical microscope image of a sample during laser pulse application and schematic of the electrode configuration for two-probe resistance measurements. (\textbf{C}) Temperature profile estimated by finite-element simulations. The inset shows the spatial temperature profile during the laser pulse. (\textbf{D} and \textbf{E}) Temperature dependences of the two-probe resistance of V$_{1-x}$W$_x$O$_2$ for $x=0.054$ (\textbf{D}) and $0.117$ (\textbf{E}). Laser quenching at low temperature reduces the resistance by more than seven orders of magnitude.
	}
	\label{fig:quenching} 
\end{figure}

To experimentally implement thermal quenching, we exploited rapid thermal diffusion following laser-induced local heating. For two-terminal resistance measurements, electrodes were patterned on a V$_{1-x}$W$_x$O$_2$ thin film grown on a TiO$_2$ substrate. A pulsed laser (fluence, $\sim$10 mJ cm$^{-2}$; wavelength, 532 nm; pulse duration, 5 ns) was focused onto the electrode gap (Fig.~\ref{fig:quenching}B), while the resistance between them was monitored with a bias voltage applied. Although the fluence is comparable to that used in previous ultrafast photoinduced-transition studies on VO$_2$ \cite{hilton2007enhanced}, the pulse duration is longer by approximately five orders of magnitude, resulting in a much lower peak power density, which is expected to be insufficient to drive an ultrafast photoinduced transition. The incident light causes a rapid, localized temperature rise in the film, while the substrate temperature $T_{\rm{o}}$ was maintained below the metal-insulator transition temperature $T_{\rm{MI}}$. After laser irradiation, the large thermal gradient drives thermal diffusion, leading to rapid cooling. Finite-element simulations indicate a temperature rise exceeding 400 K in the film, followed by ultrafast cooling occurring within approximately 100 ns (Fig.~\ref{fig:quenching}C and Fig.~\ref{fig:sup_fem}).

We first performed pulsed-laser irradiation experiments on nondoped VO$_2$. A single pulse reduced the resistance from an initial value greater than 10 G$\Omega$ to 20 k$\Omega$, but the resistance rapidly increased to over 1 M$\Omega$ within 100 ns (Fig.~\ref{fig:sup_vo2}). This timescale matches the simulated thermal relaxation, showing that the metal–insulator transition completes during the rapid cooling, as in a conventional temperature sweep. Thus, even at a cooling rate as high as 10$^9$ K s$^{-1}$, the supercooled metallic phase does not persist on laboratory timescales ($\sim$1 s) in nondoped VO$_2$. This indicates that slower kinetics of the metal–insulator transition are necessary for stabilizing a supercooled metallic phase.

We then investigated the phase-competition region of the V$_{1-x}$W$_x$O$_2$ system ($x = 0.05$–$0.15$), where sluggish transformation kinetics are expected due to competing phases. On the low-doping side ($x < 0.07$), the insulating phase is continuously connected to that of nondoped VO$_2$, characterized by V–V dimerization. In contrast, on the high-doping side ($x > 0.11$), the insulating phase evolves toward the $x = 0.33$ composition, where insulating behavior is associated with a W$^{6+}$–V$^{3+}$–V$^{3+}$ triple-period structure \cite{israelsson1970phase, pourroy1983orbitally}. We performed pulsed-laser quenching experiments at $x = 0.054$ (low-doping side) and $x = 0.117$ (high-doping side). At $x = 0.054$, the metal–insulator transition is observed below 100 K under slow cooling ($5 \times 10^{-2}$ K s$^{-1}$), with the resistance exceeding 100 M$\Omega$ below 30 K (Fig.~\ref{fig:quenching}D). Upon laser pulse irradiation at $T_{\rm{o}} = 5$ K, the post-pulse resistance dramatically drops to a few hundred ohms, matching that of the high-temperature metallic phase. Subsequent slow heating reverts the quenched metallic state to the insulating phase, and further heating induces the equilibrium insulator-to–metal transition at $T_{\rm{MI}}$, indicating that the quenched metallic phase relaxes back to equilibrium above 60 K. Analogous experiments at $x = 0.117$ also reveal a metastable metallic phase hidden behind the equilibrium insulating state, despite the distinct insulating mechanism from the low-doping regime (Fig.~\ref{fig:quenching}E). Thus, thermal quenching stabilizes a metastable metallic phase in the non-triangular system V$_{1-x}$W$_x$O$_2$ when tuned into a phase-competition regime.

\subsection*{Temperature–composition dependence of transformation kinetics}

To elucidate the conditions under which a thermally quenched metastable metallic phase emerges, we investigated the post-quench transformation kinetics as a function of $T_{\rm{o}}$ and $x$. After laser quenching at selected $T_{\rm{o}}$, isothermal time evolutions of resistance were measured for $x$ = 0.054, 0.069, 0.071, and 0.117 (Fig.~\ref{fig:relaxation}A–D). In all cases, the resistance increased with time, indicating relaxation toward the equilibrium insulating phase. By overlaying the time evolution of post-quench resistance onto the temperature-dependent resistance curves (Fig.~\ref{fig:relaxation}E–H), we directly compared the resistance windows for isothermal transformations and temperature sweeps. For $x = 0.069$ and $0.071$, the resistance measured $10^{4}$ s after quenching exceeds that obtained during temperature sweeps at a sweep rate of 5 $\times$ 10$^{-2}$ K s$^{-1}$, indicating that the transformation kinetics are slower than the temperature-sweep timescale. Notably, for $x = 0.071$, the resistance increases only by a factor of two upon cooling through the phase transition, giving the appearance of a metallic ground state. However, the persistent increase in resistance over time and its strong dependence on cooling rate (Fig.~\ref{fig:sup_ratedep}) indicate that this apparent metallicity is due to incomplete relaxation. Thus, these results suggest that thermally quenched metastable metallic phases are observed over the composition range $x = 0.054$--$0.117$.

Time–temperature–transformation (TTT) diagrams were constructed based on these data to systematically map transformation kinetics. The insulating-phase fraction, $f_i$, was estimated from the measured resistances via the general effective medium (GEM) equation \cite{mclachlan1990electrical, shibuya2011x}, thus converting the isothermal resistance curves to phase-fraction evolutions (Fig.~\ref{fig:relaxation}I–L). These TTT diagrams reveal that the transformation kinetics can vary over nearly ten orders of magnitude depending on temperature and composition, with extremely slow transitions at low temperatures. We define the metastable lifetime $\tau$ as the time for $f_i$ to reach 0.5. A model that accounts for both the free-energy driving force and a thermally activated barrier, as applied to BaV$_{10-x}$Ti$_x$O$_{15}$ \cite{katsufuji2020nucleation}, captures the non-monotonic temperature dependence of $\tau$ (Fig.~\ref{fig:sup_ttt}). At low temperatures (below 50~K), this model reduces to a simple Arrhenius-type dependence, $\tau = \tau_0 \exp(\Delta / k_B T)$, allowing estimation of the activation barrier: $\Delta$ of approximately 520~K for $x = 0.054$, 560~K for $x = 0.069$, and 650~K for $x = 0.117$. Thus, the metastable lifetime is governed by thermally activated kinetics with an energy barrier on the order of several tens of meV.

\subsection*{Optical microscopy of transformation kinetics}

\begin{figure}
	\centering
	\includegraphics[width=0.9\textwidth]{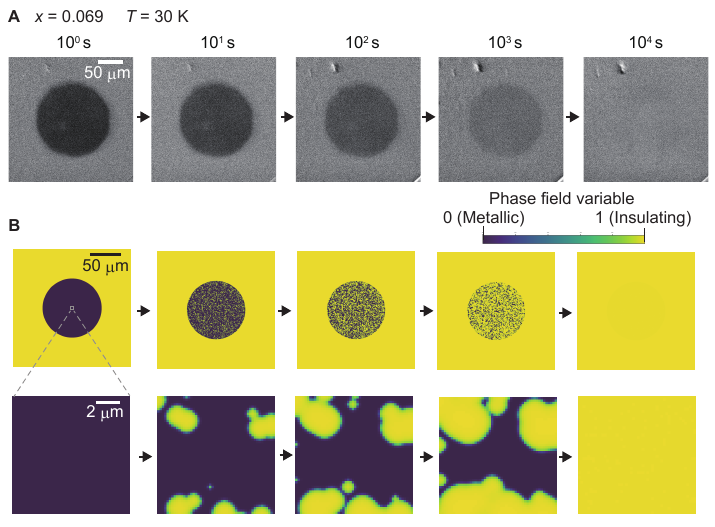}
	\caption{\textbf{Real-space imaging of the metal-insulator transition.}
		(\textbf{A}) Isothermal time evolution of optical microscope images at 30 K after pulsed laser irradiation, where the time origin is defined by the laser pulse. The dark circular regions indicate reduced reflectance due to the formation of a metastable metallic phase by thermal quenching. (\textbf{B}) Phase-field simulation of the spatial evolution of the phase transformation at two different length scales. Snapshots of the entire simulation area (200~$\mu$m $\times$ 200~$\mu$m) show that the circular domain appears to evolve nearly uniformly, whereas magnified views (5~$\mu$m $\times$ 5~$\mu$m) reveal spatial inhomogeneity consisting of a mixture of small domains formed through repeated nucleation events.}
	\label{fig:microscopy}
\end{figure}

To identify the characteristic length scale associated with the activation barrier, we performed real-space optical imaging of the isothermal relaxation following laser quenching for $x = 0.069$. Immediately after laser irradiation, a dark region appears that matches the laser spot, indicating the successful creation of the metastable metallic phase (Fig.~\ref{fig:microscopy}A). As time elapses, the contrast between the metallic and insulating regions gradually weakens while maintaining the circular footprint. This indicates that the metal–insulator transition proceeds locally at each micrometer-scale position, rather than via lateral motion of a well-defined phase boundary. These results suggest that the transformation kinetics are spatially homogeneous on the micrometer scale, implying that the activation barrier is determined by processes occurring at sub-micrometer length scales.

To infer the sub-micrometer-scale dynamics, we performed phase-field simulations incorporating nucleation and growth of insulating domains. The spatial morphology during the relaxation of a metastable domain depends on the interfacial energy between the metastable and stable phases: larger interfacial energy leads to growth-dominated kinetics characterized by boundary propagation, whereas smaller interfacial energy results in transformation patterns that appear spatially homogeneous on the micrometer scale (Fig.~\ref{fig:sup_pf}). The latter reproduces the experimental observations. Closer inspection of this regime indicates that the transformation proceeds via successive nucleation events and completes when domain growth bridges the spacing between nuclei, corresponding to nucleation-dominated kinetics (Fig.~\ref{fig:microscopy}B). The apparent spatial homogeneity on the micrometer scale thus indicates that multiple nucleation events occur before domains grow over micrometer distances. Based on insights from these simulations, we make an order-of-magnitude estimate for the growth velocity in the experiments to be $\lesssim 10^{-6}$ m s$^{-1}$ when $\tau$ exceeds 1 s at low temperatures. Such slow domain growth is characteristic of a creep regime \cite{matsuura2023low}, indicating that the activation barrier is associated with a pinning mechanism that limits domain-wall motion.

\subsection*{Bond-breaking process as an origin of metastability}

\begin{figure}
	\centering
	\includegraphics[width=0.6\textwidth]{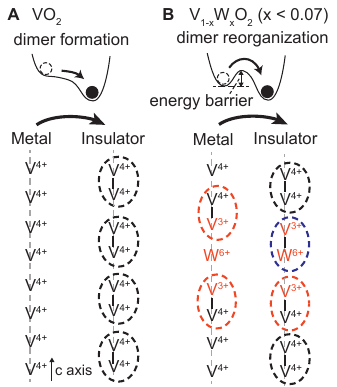}
	\caption{\textbf{Evolution of dimer configurations across the metal–insulator transition.}
		(\textbf{A}) Dimer configurations in nondoped VO$_2$. V ions are uniformly arranged along the $c$ axis in the metallic phase, whereas V–V dimers form in the insulating phase. The dimerization is considered to occur without an activation barrier within the harmonic phonon picture. (\textbf{B}) Dimer configurations in V$_{1-x}$W$_x$O$_2$ with $x < 0.07$. Local V–V dimers are formed adjacent to W sites already in the metallic phase  \cite{netsianda2008displacive,ling2019w,li2025study}. Upon the transition to the insulating phase, one of these dimers is broken to establish the long-period superstructure. We propose that this bond rearrangement contributes to the activation barrier.}
	\label{fig:concept}
\end{figure}

To discuss the pinning mechanism of domain growth, we focus on V–V dimer bonding. The metal–insulator transition in VO$_2$ is understood in terms of the formation of V$_2$O$_{10}$ molecular units associated with V–V dimerization \cite{hiroi2015structural} (Fig.~\ref{fig:concept}A). Harmonic phonon calculations for metallic rutile VO$_2$ reveal dynamical instabilities of the high-symmetry metallic structure, while anharmonic effects stabilize the metallic phase at high temperatures \cite{budai2014metallization}. The dimerization-related instability is sensitive to the Hubbard $U$; for finite $U$, the phonon frequency of the V-dimerization mode along the rutile $c$ axis becomes imaginary \cite{kim2013correlation}. In our calculations, this lattice instability persists when W substitution is incorporated within the virtual crystal approximation (VCA) (Fig.~\ref{fig:sup_phonon}), which treats elemental substitution as a spatially uniform modification of the electronic structure and therefore does not capture local structural variations around W dopants . The inability of the VCA to capture the metastability of the metallic phase suggests that the pinning mechanism is associated with local structural variations introduced by W substitution.

The local structural variation induced by partial substitution of V$^{4+}$ ions with W$^{6+}$ ions has been investigated by X-ray absorption fine-structure (XAFS) measurements \cite{tang1985local, tan2012unraveling} and first-principles calculations \cite{netsianda2008displacive, ling2019w, li2025study}. In the metallic phase, the nearest-neighbor W$^{6+}$–V$^{3+}$ distance along the $c$ axis is elongated by approximately 5\% \cite{tang1985local, tan2012unraveling}, while the adjacent V$^{3+}$–V$^{4+}$ distance is correspondingly shortened by about 8\% \cite{netsianda2008displacive, ling2019w, li2025study}, resulting in the formation of V$^{3+}$–V$^{4+}$ dimers on both sides of the W$^{6+}$ ion, as illustrated in Fig.~\ref{fig:concept}B. In the insulating phase, only one of these local dimers is incorporated into the periodic dimer arrangement of the superstructure (Fig.~\ref{fig:concept}B) \cite{tan2012unraveling, ling2019w, li2025study}. Therefore, the metal-insulator transition involves breaking of the other pre-existing V$^{3+}$–V$^{4+}$ dimer. We propose that the activation barrier in the low-doping region ($x < 0.07$) may originate from this bond-breaking process. In the high-doping regime ($x > 0.11$), no clear symmetry lowering is observed across the metal–insulator transition, and the origin of the insulating state remains unresolved \cite{okuyama2015x}. The $c$-axis lattice parameter elongates by about 1\% across the transition \cite{okuyama2015x}, suggesting local structural rearrangements. Clarifying the microscopic mechanism in this regime is therefore an important subject for future investigation.

Microscopic origins of metastability in electronic systems have often been attributed to geometrical frustration \cite{kagawa2013charge, stojchevska2014ultrafast} and magnetic topology \cite{oike2016interplay}. The present results suggest that elemental substitution provides an alternative route to metastable metallic phases by creating an activation barrier through local structural variations. In the low-doping regime of V$_{1-x}$W$_x$O$_2$, this barrier may originate from the breaking of pre-existing V--V dimers, whereas the microscopic origin in the high-doping regime remains unresolved. Such a bond-breaking perspective may apply to the geometrically frustrated system 1T-TaS$_2$, where the nearly commensurate--to--commensurate transition involves breaking and reforming Ta--Ta bonds in Star-of-David clusters \cite{stojchevska2014ultrafast}, and the high-temperature phase can be metastabilized by rapid quenching \cite{yoshida2014controlling}. This comparison suggests that elemental substitution can create metastable metallic phases in non-triangular systems through mechanisms analogous to those in triangular-lattice systems. Understanding how local structural rearrangements create activation barriers is therefore important for developing microscopic design principles for thermally quenched metastable phases.

Thermal and optical excitations offer complementary methods for examining the formation of metastable electronic phases. In the present study, we employed thermal quenching, a method widely used in metallurgy, to access such phases. Metastable electronic phases have also been induced by photoexcitation \cite{stojchevska2014ultrafast}, which generates nonequilibrium electronic distributions distinct from those produced by thermal quenching. Clarifying how these different excitations affect local structural rearrangements, including changes in bonding configurations, may provide a common framework for describing metastability.


\begin{figure}
		\centering
		\includegraphics[width=0.9\textwidth]{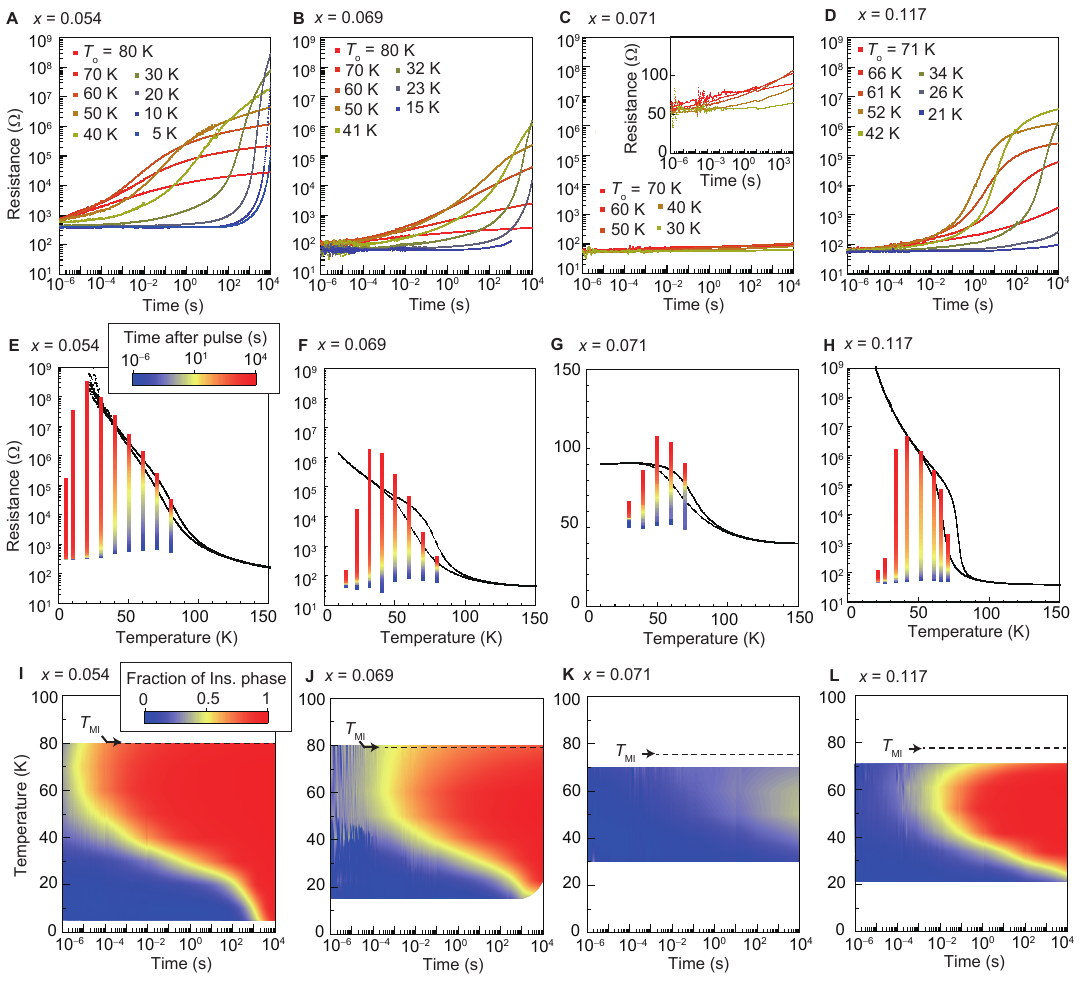}
		\caption{\textbf{Isothermal time evolutions from metastable metallic phase to the most stable insulating phase.}
			(\textbf{A} to \textbf{D}) Isothermal time evolution of the resistance of
			V$_{1-x}$W$_x$O$_2$ for $x=0.054$ (\textbf{A}), $0.069$ (\textbf{B}),
			$0.071$ (\textbf{C}), and $0.117$ (\textbf{D}) after laser-pulse application,
			with the pulse time defined as $t=0$. The substrate temperature $T_{\mathrm{o}}$ was held constant during each measurement. (\textbf{E} to \textbf{H}) Temperature dependence of the resistance measured at a sweep rate of $5 \times 10^{-2}$ K s$^{-1}$ for $x=0.054$ (\textbf{E}), $0.069$ (\textbf{F}), $0.071$ (\textbf{G}), and $0.117$ (\textbf{H}). The post-quench time evolution of the resistance is overlaid as color maps. (\textbf{I} to \textbf{L}) Time-temperature-transformation diagrams for
			$x=0.054$ (\textbf{I}), $0.069$ (\textbf{J}), $0.071$ (\textbf{K}),
			and $0.117$ (\textbf{L}). The insulating-phase fraction was estimated from the resistance data using the general effective-medium equation  \cite{mclachlan1990electrical,shibuya2011x}.}
		\label{fig:relaxation}
	\end{figure}

\clearpage 

%
\bibliography{Oike_ref} 

@article{angell1995formation,
  title={Formation of glasses from liquids and biopolymers},
  author={Angell, C Austen},
  journal={Science},
  volume={267},
  number={5206},
  pages={1924--1935},
  year={1995},
  publisher={American Association for the Advancement of Science}
}

@article{debenedetti2001supercooled,
  title={Supercooled liquids and the glass transition},
  author={Debenedetti, Pablo G and Stillinger, Frank H},
  journal={Nature},
  volume={410},
  number={6825},
  pages={259--267},
  year={2001},
  publisher={Nature Publishing Group UK London}
}

@article{kagawa2013charge,
  title={Charge-cluster glass in an organic conductor},
  author={Kagawa, F and Sato, T and Miyagawa, K and Kanoda, K and Tokura, Y and Kobayashi, K and Kumai, R and Murakami, Y},
  journal={Nature Physics},
  volume={9},
  number={7},
  pages={419--422},
  year={2013},
  publisher={Nature Publishing Group UK London}
}

@article{oike2015phase,
  title={Phase-change memory function of correlated electrons in organic conductors},
  author={Oike, H and Kagawa, F and Ogawa, N and Ueda, A and Mori, H and Kawasaki, M and Tokura, Y},
  journal={Physical Review B},
  volume={91},
  number={4},
  pages={041101},
  year={2015},
  publisher={APS}
}

@article{sato2017electronic,
  title={Electronic crystal growth},
  author={Sato, Takuro and Miyagawa, K and Kanoda, K},
  journal={Science},
  volume={357},
  number={6358},
  pages={1378--1381},
  year={2017},
  publisher={American Association for the Advancement of Science}
}

@article{sasaki2017crystallization,
  title={Crystallization and vitrification of electrons in a glass-forming charge liquid},
  author={Sasaki, S and Hashimoto, K and Kobayashi, Ryuji and Itoh, K and Iguchi, S and Nishio, Y and Ikemoto, Y and Moriwaki, T and Yoneyama, N and Watanabe, M and others},
  journal={Science},
  volume={357},
  number={6358},
  pages={1381--1385},
  year={2017},
  publisher={American Association for the Advancement of Science}
}

@article{yoshida2014controlling,
  title={Controlling charge-density-wave states in nano-thick crystals of {1T-TaS$_2$}},
  author={Yoshida, Masaro and Zhang, Yijin and Ye, Jianting and Suzuki, Ryuji and Imai, Yasuhiko and Kimura, Shigeru and Fujiwara, Akihiko and Iwasa, Yoshihiro},
  journal={Scientific Reports},
  volume={4},
  number={1},
  pages={7302},
  year={2014},
  publisher={Nature Publishing Group UK London}
}

@article{oike2016interplay,
  title={Interplay between Topological and Thermodynamic Stability in a Metastable Magnetic Skyrmion Lattice},
  author={Oike, Hiroshi and Kikkawa, Akiko and Kanazawa, Naoya and Taguchi, Yasujiro and Kawasaki, Masashi and Tokura, Yoshinori and Kagawa, Fumitaka},
  journal={Nature Physics},
  volume={12},
  number={1},
  pages={62--66},
  year={2016},
  publisher={Nature Publishing Group}
}

@article{karube2016robust,
  title={Robust metastable skyrmions and their triangular--square lattice structural transition in a high-temperature chiral magnet},
  author={Karube, K and White, JS and Reynolds, N and Gavilano, JL and Oike, H and Kikkawa, A and Kagawa, F and Tokunaga, Y and R{\o}nnow, Henrik M and Tokura, Y and others},
  journal={Nature Materials},
  volume={15},
  number={12},
  pages={1237--1242},
  year={2016},
  publisher={Nature Publishing Group UK London}
}

@article{berruto2018laser,
  title={Laser-induced skyrmion writing and erasing in an ultrafast cryo-Lorentz transmission electron microscope},
  author={Berruto, G and Madan, I and Murooka, Y and Vanacore, GM and Pomarico, E and Rajeswari, J and Lamb, R and Huang, P and Kruchkov, AJ and Togawa, Y and others},
  journal={Physical Review Letters},
  volume={120},
  number={11},
  pages={117201},
  year={2018},
  publisher={APS}
}

@article{birch2019increased,
  title={Increased lifetime of metastable skyrmions by controlled doping},
  author={Birch, MT and Takagi, R and Seki, S and Wilson, MN and Kagawa, F and {\v{S}}tefan{\v{c}}i{\v{c}}, Ale{\v{s}} and Balakrishnan, Geetha and Fan, R and Steadman, P and Ottley, CJ and others},
  journal={Physical Review B},
  volume={100},
  number={1},
  pages={014425},
  year={2019},
  publisher={APS}
}

@article{matsuura2021kinetic,
  title={Kinetic pathway facilitated by a phase competition to achieve a metastable electronic phase},
  author={Matsuura, Keisuke and Oike, Hiroshi and Kocsis, Vilmos and Sato, Takuro and Tomioka, Yasuhide and Kaneko, Yoshio and Nakamura, Masao and Taguchi, Yasujiro and Kawasaki, Masashi and Tokura, Yoshinori and others},
  journal={Physical Review B},
  volume={103},
  number={4},
  pages={L041106},
  year={2021},
  publisher={APS}
}

@article{matsuura2023thermodynamic,
  title={Thermodynamic determination of the equilibrium first-order phase-transition line hidden by hysteresis in a phase diagram},
  author={Matsuura, Keisuke and Nishizawa, Yo and Kriener, Markus and Kurumaji, Takashi and Oike, Hiroshi and Tokura, Yoshinori and Kagawa, Fumitaka},
  journal={Scientific Reports},
  volume={13},
  number={1},
  pages={6876},
  year={2023},
  publisher={Nature Publishing Group UK London}
}

@article{katsufuji2020nucleation,
  title={Nucleation and growth of orbital ordering},
  author={Katsufuji, Takuro and Kajita, Tomomasa and Yano, Suguru and Katayama, Yumiko and Ueno, Kazunori},
  journal={Nature Communications},
  volume={11},
  number={1},
  pages={2324},
  year={2020},
  publisher={Nature Publishing Group UK London}
}

@article{oike2018kinetic,
  title={Kinetic approach to superconductivity hidden behind a competing order},
  author={Oike, Hiroshi and Kamitani, Manabu and Tokura, Yoshinori and Kagawa, Fumitaka},
  journal={Science Advances},
  volume={4},
  number={10},
  pages={eaau3489},
  year={2018},
  publisher={American Association for the Advancement of Science}
}

@article{mori1998systematic,
  title={Systematic study of the electronic state in $\theta$-type {BEDT-TTF} organic conductors by changing the electronic correlation},
  author={Mori, Hatsumi and Tanaka, Shoji and Mori, Takehiko},
  journal={Physical Review B},
  volume={57},
  number={19},
  pages={12023},
  year={1998},
  publisher={APS}
}

@article{wilson1975charge,
  title={Charge-density waves and superlattices in the metallic layered transition metal dichalcogenides},
  author={Wilson, Jl A and Di Salvo, FJ and Mahajan, S},
  journal={Advances in Physics},
  volume={24},
  number={2},
  pages={117--201},
  year={1975},
  publisher={Taylor \& Francis}
}

@article{stojchevska2014ultrafast,
  title={Ultrafast switching to a stable hidden quantum state in an electronic crystal},
  author={Stojchevska, L and Vaskivskyi, I and Mertelj, T and Kusar, P and Svetin, D and Brazovskii, S and Mihailovic, D},
  journal={Science},
  volume={344},
  number={6180},
  pages={177--180},
  year={2014},
  publisher={American Association for the Advancement of Science}
}

@article{vaskivskyi2015controlling,
  title={Controlling the metal-to-insulator relaxation of the metastable hidden quantum state in {1T-TaS$_2$}},
  author={Vaskivskyi, Igor and Gospodaric, Jan and Brazovskii, Serguei and Svetin, Damjan and Sutar, Petra and Goreshnik, Evgeny and Mihailovic, Ian A and Mertelj, Tomaz and Mihailovic, Dragan},
  journal={Science Advances},
  volume={1},
  number={6},
  pages={e1500168},
  year={2015},
  publisher={American Association for the Advancement of Science}
}

@article{shibuya2010metal,
  title={Metal-insulator transition in epitaxial {V$_{1-x}$W$_x$O$_2$} ($0\leq x\leq 0.33$) thin films},
  author={Shibuya, Keisuke and Kawasaki, Masashi and Tokura, Yoshinori},
  journal={Applied Physics Letters},
  volume={96},
  number={2},
  year={2010},
  publisher={AIP Publishing}
}

@article{oike2025thermally,
  title={Thermally quenched metastable phase in the Ising model with competing interactions},
  author={Oike, Hiroshi and Suwa, Hidemaro and Takahashi, Yasunori and Kagawa, Fumitaka},
  journal={Physical Review B},
  volume={112},
  number={6},
  pages={064409},
  year={2025},
  publisher={APS}
}

@article{hilton2007enhanced,
  title={Enhanced photosusceptibility near ${T}_c$ for the light-induced insulator-to-metal phase transition in vanadium dioxide},
  author={Hilton, DJ and Prasankumar, RP and Fourmaux, S and Cavalleri, A and Brassard, D and El Khakani, MA and Kieffer, JC and Taylor, AJ and Averitt, RD},
  journal={Physical Review Letters},
  volume={99},
  number={22},
  pages={226401},
  year={2007},
  publisher={APS}
}

@article{morrison2014photoinduced,
  title={A photoinduced metal-like phase of monoclinic {VO}$_2$ revealed by ultrafast electron diffraction},
  author={Morrison, Vance R and Chatelain, Robert P and Tiwari, Kunal L and Hendaoui, Ali and Bruh{\'a}cs, Andrew and Chaker, Mohamed and Siwick, Bradley J},
  journal={Science},
  volume={346},
  number={6208},
  pages={445--448},
  year={2014},
  publisher={American Association for the Advancement of Science}
}

@article{shibuya2011x,
  title={X-ray induced insulator-metal transition in a thin film of electron-doped {VO}$_2$},
  author={Shibuya, K and Okuyama, D and Kumai, R and Yamasaki, Y and Nakao, H and Murakami, Y and Taguchi, Y and Arima, T and Kawasaki, M and Tokura, Y},
  journal={Physical Review B},
  volume={84},
  number={16},
  pages={165108},
  year={2011},
  publisher={APS}
}

@article{okuyama2015x,
  title={X-ray study of metal-insulator transitions induced by {W} doping and photoirradiation in {VO}$_2$ films},
  author={Okuyama, D and Shibuya, K and Kumai, R and Suzuki, T and Yamasaki, Y and Nakao, H and Murakami, Y and Kawasaki, M and Taguchi, Y and Tokura, Y and others},
  journal={Physical Review B},
  volume={91},
  number={6},
  pages={064101},
  year={2015},
  publisher={APS}
}

@article{longo1970vo2,
  title   = {A Refinement of the Structure of {VO}$_2$},
  author  = {Longo, John M. and Kierkegaard, Peder},
  journal = {Acta Chemica Scandinavica},
  volume  = {24},
  pages   = {420--426},
  year    = {1970}
}

@article{israelsson1970phase,
  title={The phase relations in the {VO$_2$| WO$_2$} system},
  author={Israelsson, Mats and Kihlborg, Lars},
  journal={Materials Research Bulletin},
  volume={5},
  number={1},
  pages={19--29},
  year={1970},
  publisher={Elsevier}
}

@article{pourroy1983orbitally,
  title={The orbitally degenerate binuclear unit t$^2_{2g}$-t$^2_{2g}$: Part II: Interaction between vanadium III ions in the {WV$_2$O$_6$} compound},
  author={Pourroy, G and Drillon, M and Padel, L and Bernier, JC},
  journal={Physica B+C},
  volume={123},
  number={1},
  pages={21--26},
  year={1983},
  publisher={Elsevier}
}

@article{mclachlan1990electrical,
  title={Electrical resistivity of composites},
  author={McLachlan, David S and Blaszkiewicz, Michael and Newnham, Robert E},
  journal={Journal of the American Ceramic Society},
  volume={73},
  number={8},
  pages={2187--2203},
  year={1990},
  publisher={Wiley Online Library}
}

@article{matsuura2023low,
  title={Low-temperature hysteresis broadening emerging from domain-wall creep dynamics in a two-phase competing system},
  author={Matsuura, Keisuke and Nishizawa, Yo and Kinoshita, Yuto and Kurumaji, Takashi and Miyake, Atsushi and Oike, Hiroshi and Tokunaga, Masashi and Tokura, Yoshinori and Kagawa, Fumitaka},
  journal={Communications Materials},
  volume={4},
  number={1},
  pages={71},
  year={2023},
  publisher={Nature Publishing Group UK London}
}

@article{hiroi2015structural,
  title={Structural instability of the rutile compounds and its relevance to the metal--insulator transition of {VO}$_2$},
  author={Hiroi, Zenji},
  journal={Progress in Solid State Chemistry},
  volume={43},
  number={1-2},
  pages={47--69},
  year={2015},
  publisher={Elsevier}
}

@article{budai2014metallization,
  title={Metallization of vanadium dioxide driven by large phonon entropy},
  author={Budai, John D and Hong, Jiawang and Manley, Michael E and Specht, Eliot D and Li, Chen W and Tischler, Jonathan Z and Abernathy, Douglas L and Said, Ayman H and Leu, Bogdan M and Boatner, Lynn A and others},
  journal={Nature},
  volume={515},
  number={7528},
  pages={535--539},
  year={2014},
  publisher={Nature Publishing Group UK London}
}

@article{kim2013correlation,
  title = {Correlation-assisted phonon softening and the orbital-selective Peierls transition in {VO}$_2$},
  author = {Kim, Sooran and Kim, Kyoo and Kang, Chang-Jong and Min, B. I.},
  journal = {Phys. Rev. B},
  volume = {87},
  issue = {19},
  pages = {195106},
  numpages = {5},
  year = {2013},
  month = {May},
  publisher = {American Physical Society},
  }

@article{tang1985local,
  title={Local atomic and electronic arrangements in {W$_x$V$_{1-x}$O$_2$}},
  author={Tang, C and Georgopoulos, P and Fine, ME and Cohen, JB and Nygren, M and Knapp, GS and Aldred, A},
  journal={Physical Review B},
  volume={31},
  number={2},
  pages={1000},
  year={1985},
  publisher={APS}
}

@article{tan2012unraveling,
  title={Unraveling metal-insulator transition mechanism of {VO}$_2$ triggered by tungsten doping},
  author={Tan, Xiaogang and Yao, Tao and Long, Ran and Sun, Zhihu and Feng, Yajuan and Cheng, Hao and Yuan, Xun and Zhang, Wenqing and Liu, Qinghua and Wu, Changzheng and others},
  journal={Scientific Reports},
  volume={2},
  number={1},
  pages={466},
  year={2012},
  publisher={Nature Publishing Group UK London}
}

@article{netsianda2008displacive,
  title={The displacive phase transition of vanadium dioxide and the effect of doping with tungsten},
  author={Netsianda, Makondelele and Ngoepe, Phuti E and Catlow, C Richard A and Woodley, Scott M},
  journal={Chemistry of Materials},
  volume={20},
  number={5},
  pages={1764--1772},
  year={2008},
  publisher={ACS Publications}
}

@article{ling2019w,
  title={W doping and voltage driven metal--insulator transition in {VO}$_2$ nano-films for smart switching devices},
  author={Ling, Chen and Zhao, Zhengjing and Hu, Xinyuan and Li, Jingbo and Zhao, Xushan and Wang, Zongguo and Zhao, Yongjie and Jin, Haibo},
  journal={ACS Applied Nano Materials},
  volume={2},
  number={10},
  pages={6738--6746},
  year={2019},
  publisher={ACS Publications}
}

@article{li2025study,
  title={Study on the Mechanism of Phase Transition of {VO}$_2$ with Low-Concentration {W} Doping},
  author={Li, Xiaoyue and Wang, Yixin and Li, Hong and Shi, Sujun and Zheng, Bing},
  journal={Journal of Electronic Materials},
  volume={54},
  number={12},
  pages={11345--11355},
  year={2025},
  publisher={Springer}
}

@article{togo2023first,
  title = {First-Principles Phonon Calculations with {Phonopy} and {Phono3py}},
  author = {Togo, Atsushi},
  journal = {Journal of the Physical Society of Japan},
  volume = {92},
  number = {1},
  pages = {012001},
  year = {2023}
}

@article{kresse1996efficient,
  title = {Efficient iterative schemes for ab initio total-energy calculations using a plane-wave basis set},
  author = {Kresse, G. and Furthm{\"u}ller, J.},
  journal = {Physical Review B},
  volume = {54},
  number = {16},
  pages = {11169--11186},
  year = {1996}
}

@article{perdew1996generalized,
  title={Generalized gradient approximation made simple},
  author={Perdew, John P and Burke, Kieron and Ernzerhof, Matthias},
  journal={Physical Review Letters},
  volume={77},
  number={18},
  pages={3865},
  year={1996},
  publisher={APS}
}

@article{kresse1999ultrasoft,
  title = {From ultrasoft pseudopotentials to the projector augmented-wave method},
  author = {Kresse, G. and Joubert, D.},
  journal = {Physical Review B},
  volume = {59},
  pages = {1758--1775},
  year = {1999}
}

@article{dudarev1998electron,
  title = {Electron-energy-loss spectra and the structural stability of nickel oxide: An {LSDA}+{U} study},
  author = {Dudarev, S. L. and Botton, G. A. and Savrasov, S. Y. and Humphreys, C. J. and Sutton, A. P.},
  journal = {Physical Review B},
  volume = {57},
  pages = {1505--1509},
  year = {1998}
}

@article{momma2011vesta,
  title = {{VESTA} 3 for three-dimensional visualization of crystal, volumetric and morphology data},
  author = {Momma, Koichi and Izumi, Fujio},
  journal = {Journal of Applied Crystallography},
  volume = {44},
  number = {6},
  pages = {1272--1276},
  year = {2011}
}

@article{slichter1972pressure,
  title={Pressure-induced electronic changes in compounds of iron},
  author={Slichter, CP and Drickamer, HG},
  journal={The Journal of Chemical Physics},
  volume={56},
  number={5},
  pages={2142--2160},
  year={1972},
  publisher={American Institute of Physics}
}

@article{tokoro2015external,
  title={External stimulation-controllable heat-storage ceramics},
  author={Tokoro, Hiroko and Yoshikiyo, Marie and Imoto, Kenta and Namai, Asuka and Nasu, Tomomichi and Nakagawa, Kosuke and Ozaki, Noriaki and Hakoe, Fumiyoshi and Tanaka, Kenji and Chiba, Kouji and others},
  journal={Nature communications},
  volume={6},
  number={1},
  pages={7037},
  year={2015},
  publisher={Nature Publishing Group UK London}
}

@article{uchimura2020robust,
  title={A robust thermal-energy-storage property associated with electronic phase transitions for quadruple perovskite oxides},
  author={Uchimura, Tasuku and Yamada, Ikuya},
  journal={Chemical Communications},
  volume={56},
  number={41},
  pages={5500--5503},
  year={2020},
  publisher={The Royal Society of Chemistry}
}

@article{tsai2004effect,
  title={Effect of grain curvature on nano-indentation measurements of thin films},
  author={Tsai, Kuang-Yue and Chin, Tsung-Shune and Shieh, Han-Ping D},
  journal={Japanese journal of applied physics},
  volume={43},
  number={9R},
  pages={6268},
  year={2004}
}

@article{oike2021real,
  title={Real-Space Observation of Emergent Complexity of Phase Evolution in Micrometer-Sized {IrTe}$_2$ Crystals},
  author={Oike, H and Takeda, K and Kamitani, M and Tokura, Y and Kagawa, F},
  journal={Physical Review Letters},
  volume={127},
  number={14},
  pages={145701},
  year={2021},
  publisher={APS}
}

@article{li2015synthesis,
  title={Synthesis of monoclinic {IrTe}$_2$ under high pressure and its physical properties},
  author={Li, Xiaoqun and Yan, J-Q and Singh, David J and Goodenough, JB and Zhou, J-S},
  journal={Physical Review B},
  volume={92},
  number={15},
  pages={155118},
  year={2015},
  publisher={APS}
}

@article{ko2015charge,
  title={Charge-ordering cascade with spin--orbit Mott dimer states in metallic iridium ditelluride},
  author={Ko, K-T and Lee, H-H and Kim, D-H and Yang, J-J and Cheong, S-W and Eom, MJ and Kim, JS and Gammag, R and Kim, K-S and Kim, H-S and others},
  journal={Nature communications},
  volume={6},
  number={1},
  pages={7342},
  year={2015},
  publisher={Nature Publishing Group UK London}
}
\bibliographystyle{sciencemag}

%
%
%
%
%
%


\section*{Acknowledgments}
H.O. thanks N. Ogawa and Y. Kohama for experimental insights, Y. Yamasaki and Y. Kozuka for simulation supports and M. Kawasaki, Y. Iwasa, Y. Tokura and H. Kageyama for valuable discussions. 
\paragraph*{Funding:}
H.~O. was funded by JST PRESTO (Grant No. JPMJPR21Q2), JSPS KAKENHI (Grant Nos. JP22H01164, JP23K22435, JP23H04861) and JSPS Core-to-Core Program (Grant No JPJSCCA20240001). M.~H was funded by JST PRESTO (Grant No. JPMJPR21Q6). H.~T. was funded by the JST FOREST Program (JPMJFR213Q). F.~K. was funded by the JSPS KAKENHI (Grant No. JP26H00383). MANA is supported by World Premier International Research Center Initiative (WPI), MEXT, Japan. 
\paragraph*{Author contributions:}
H.~O. and F.~K. conceived the project. H.~O. and Y.~T. performed the experiments and analyzed the data. K.~S. synthesized the W-doped VO$_2$ samples. M.~N. synthesized the VO$_2$ samples. H.~O. performed the phase field simulations. H.~T. and H.~O. performed the Slichter--Drickamer calculations. T.~H., M.~H. and H.~O. performed first-principles-based phonon calculations. H.~O. wrote the manuscript with critical input and revision from F.~K., and with comments from all other authors.
\paragraph*{Competing interests:}
There are no competing interests to declare.
\paragraph*{Data, code and materials availability:}
The source data and custom Python codes supporting the findings of this study have been deposited in Zenodo (DOI: 10.5281/zenodo.21486350) and will be made publicly available upon publication.


\subsection*{Supplementary materials}
Materials and Methods\\
Supplementary Text\\
Figs. S1 to S9\\
Table S1\\
References \textit{(40-\arabic{enumiv})}\\ 


\newpage


\renewcommand{\thefigure}{S\arabic{figure}}
\renewcommand{\thetable}{S\arabic{table}}
\renewcommand{\theequation}{S\arabic{equation}}
\renewcommand{\thepage}{S\arabic{page}}
\setcounter{figure}{0}
\setcounter{table}{0}
\setcounter{equation}{0}
\setcounter{page}{1} 


\begin{center}
\section*{Supplementary Materials for\\ \scititle}

Hiroshi~Oike$^{\ast}$,
Yasunori~Takahashi,
Keisuke~Shibuya,
Masaki~Nakano,
Tatsuki~Hanada,
Motoaki~Hirayama,
Hiroko~Tokoro,
Fumitaka~Kagawa\\ 
\small$^\ast$Corresponding author. Email: OIKE.Hiroshi@nims.go.jp\\
\end{center}

\subsubsection*{This PDF file includes:}
Materials and Methods\\
Supplementary Text\\
Figures S1 to S9\\
Table S1\\

\newpage


\subsection*{Materials and Methods}
\subsubsection*{Device fabrication}
V$_{1-x}$W$_x$O$_2$ (001) epitaxial thin films were grown on TiO$_2$ (001) single-crystal substrates (Shinkosha Ltd) by pulsed laser deposition at 300$^{\circ}$C under an oxygen pressure of 10–20 mTorr \cite{shibuya2010metal}. The film thicknesses for $x$ = 0.054, 0.069, 0.071, and 0.117 were 28, 46, 45, and 55 nm, respectively. Photolithography was first used to define the electrode pattern, followed by deposition of a 5 nm Ti adhesion layer and a 100 nm Au layer. The two-probe device geometry was then defined by argon-ion milling through a second photolithography mask, leaving the V$_{1-x}$W$_x$O$_2$ film beneath the electrodes and in the channel between them. The oxygen-deficient, conducting TiO$_2$ surface produced during ion milling was subsequently annealed to recover its insulating state.
\subsubsection*{Finite element simulation}
Finite element simulations of the temperature evolution during pulsed laser irradiation were performed using COMSOL Multiphysics. The model consists of a 40 nm V$_{1-x}$W$_x$O$_2$ thin film on a TiO$_2$ substrate with a 100 nm Au electrode. The bottom of the substrate was assumed to be in contact with a heat bath at 10 K. The laser pulse was incident from the thin-film side, and the incident light was assumed to be fully absorbed in the V$_{1-x}$W$_x$O$_2$ layer. To reduce the computational cost, the 5 nm Ti adhesion layer was neglected and the substrate thickness was set to 10 $\mu$m. The density, thermal conductivity, and specific heat were fixed at their room-temperature values for order-of-magnitude estimates.
\subsubsection*{Electrical measurements}
\begin{figure} 
		\centering
		\includegraphics[width=0.9\textwidth]{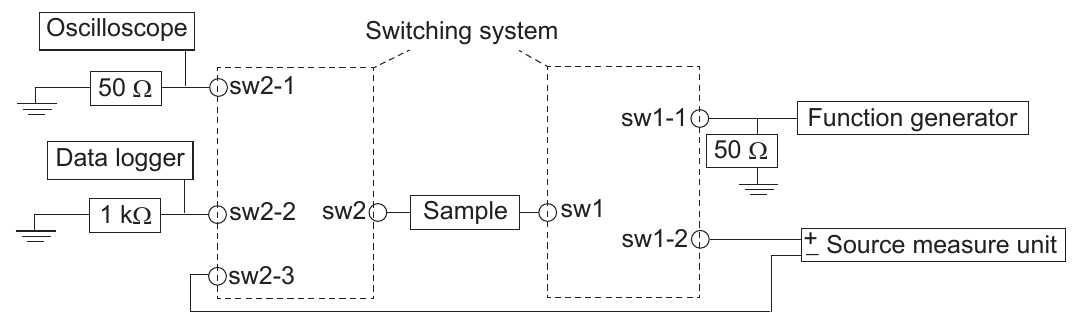} 
		\caption{\textbf{Schematic diagram of the measurement circuit.}
			SW1 and SW2 were switched using a switching system depending on the time domain. For $10^{-6}$--$10^{-4}$~s and $10^{-4}$--$10^{-2}$~s, SW1 was connected to SW1-1 and SW2 to SW2-1. For $10^{-2}$--$10^{1}$~s, SW1 was connected to SW1-1 and SW2 to SW2-2. For $10^{1}$--$10^{4}$~s, SW1 was connected to SW1-2 and SW2 to SW2-3.}
		\label{fig:sup_circuit} 
	\end{figure}

Pulsed laser irradiation was performed through a coaxial optical path of an infinity-corrected microscope onto a two-probe device. The laser source was a nanosecond Q-switched Nd:YAG laser operating at the second harmonic (Surelite, Continuum). To capture the time evolution over ten orders of magnitude, the resistance was measured in four time domains: $10^{-6}$–$10^{-4}$ s, $10^{-4}$–$10^{-2}$ s, $10^{-2}$–$10^{1}$ s, and $10^{1}$–$10^{4}$ s. A schematic of the measurement circuits is shown in Fig.~\ref{fig:sup_circuit}. In the time domains of $10^{-6}$–$10^{-4}$ s and $10^{-4}$–$10^{-2}$ s, the resistance was measured using a high-resolution oscilloscope (PXIe-5170, National Instruments), triggered by a photodiode synchronized with the laser pulse. A bias voltage was applied using a function generator (WF1947, NF Corporation), and the current was obtained from the voltage drop across a series 50~$\Omega$ resistor. The sample voltage was determined from the difference between measurements taken under positive and negative bias, thereby cancelling offset voltages. In the time domain of $10^{-2}$–$10^{1}$ s, a data logger (NI-9239, National Instruments) was used, while in $10^{1}$–$10^{4}$ s, a source measure unit (Keithley 2400) was employed. The circuit configuration for each time domain was selected before laser irradiation using a switching system (PXIe-2527, National Instruments), and each time-domain measurement was performed after a separate laser pulse. The temperature dependence of the resistance before and after laser irradiation was measured using the source measure unit.
\subsubsection*{Optical microscopy}
Time-resolved optical microscope images were acquired under continuous illumination. The illumination and detection paths were combined using a beam splitter, and the reflected light from the sample was recorded by a CMOS camera (ORCA-Flash4.0, Hamamatsu Photonics). The pulsed laser beam was introduced coaxially with the illumination path via an additional beam splitter. The timing of the laser irradiation and image acquisition was synchronized using a function generator, which provided trigger signals to both the laser and the camera. The function generator was controlled by a PC, and the interval between triggers was adjusted using the internal clock via LabVIEW. To enhance sensitivity to reflectivity changes, the grayscale values of images acquired after laser irradiation were normalized by those obtained before irradiation.

\subsubsection*{Phase fraction analysis}
The insulating volume fraction $f_i$ was estimated from the measured resistance using the general effective-medium (GEM) model:
\begin{equation}
	(1-f_i)
	\frac{\rho_l^{-1/t}-\rho^{-1/t}}
	{\rho_l^{-1/t}+A\rho^{-1/t}}
	+
	f_i
	\frac{\rho_h^{-1/t}-\rho^{-1/t}}
	{\rho_h^{-1/t}+A\rho^{-1/t}}
	=0,
	\label{eq:GEM}
\end{equation}
where
\begin{equation}
	A=\frac{1-f_c}{f_c}.
	\label{eq:GEM_A}
\end{equation}
Here, $f_c$, $t$, $\rho_l$, $\rho_h$, and $\rho$ denote the percolation threshold, critical exponent, resistivity of the metallic phase, resistivity of the insulating phase, and measured resistivity, respectively. Solving Eq.~\ref{eq:GEM} for $f_i$ gives
\begin{equation}
	f_i=
	\frac{
		\displaystyle
		\frac{1-(\rho_l/\rho)^{1/t}}
		{1+A(\rho_l/\rho)^{1/t}}
	}{
		\displaystyle
		\frac{1-(\rho_l/\rho)^{1/t}}
		{1+A(\rho_l/\rho)^{1/t}}
		-
		\frac{1-(\rho_h/\rho)^{1/t}}
		{1+A(\rho_h/\rho)^{1/t}}
	}.
	\label{eq:GEM_f_rho}
\end{equation}
Because the model depends only on resistivity ratios, the resistance $R$ can be used instead of the resistivity $\rho$, yielding
\begin{equation}
	f_i=
	\frac{
		\displaystyle
		\frac{1-(R_l/R)^{1/t}}
		{1+A(R_l/R)^{1/t}}
	}{
		\displaystyle
		\frac{1-(R_l/R)^{1/t}}
		{1+A(R_l/R)^{1/t}}
		-
		\frac{1-(R_h/R)^{1/t}}
		{1+A(R_h/R)^{1/t}}
	}.
	\label{eq:GEM_f_R}
\end{equation}
Here, $R$ is the measured resistance, and $R_l$ and $R_h$ correspond to the fully metallic ($f_i=0$) and fully insulating ($f_i=1$) states, respectively. The percolation threshold and critical exponent were set to $f_c=0.95$ and $t=1.5$, respectively, based on previous studies combining transport and structural measurements \cite{shibuya2011x}. The values of $R_l$ and $R_h$ used in the analysis are listed in Table~S1.

The value of $R_l$ was taken as the resistance measured immediately after quenching to the lowest temperature, typically 1~$\mu$s after the laser pulse, where the system is in a metallic state. Among the measured compositions, the value of $R_l$ for $x=0.054$ was approximately six to seven times larger than those for the other samples. This difference cannot be explained solely by the smaller film thickness of the $x=0.054$ sample, which was approximately half that of the others. In fact, for $x=0.054$, a finite insulating fraction was already observed at 50--80~K even 1~$\mu$s after the laser pulse, suggesting that part of the metallic phase relaxed into the insulating phase during the cooling process on the $\sim$100~ns timescale. Therefore, the value of $R_l$ may include not only the effect of film thickness but also a contribution from partial relaxation into the insulating phase during rapid cooling. Consequently, the estimated insulating fraction $f_i$ should more precisely be regarded as a relative insulating fraction defined with respect to the metallic fraction remaining approximately 1~$\mu$s after the laser pulse.

\begin{figure} 
		\centering
		\includegraphics[width=0.9\textwidth]{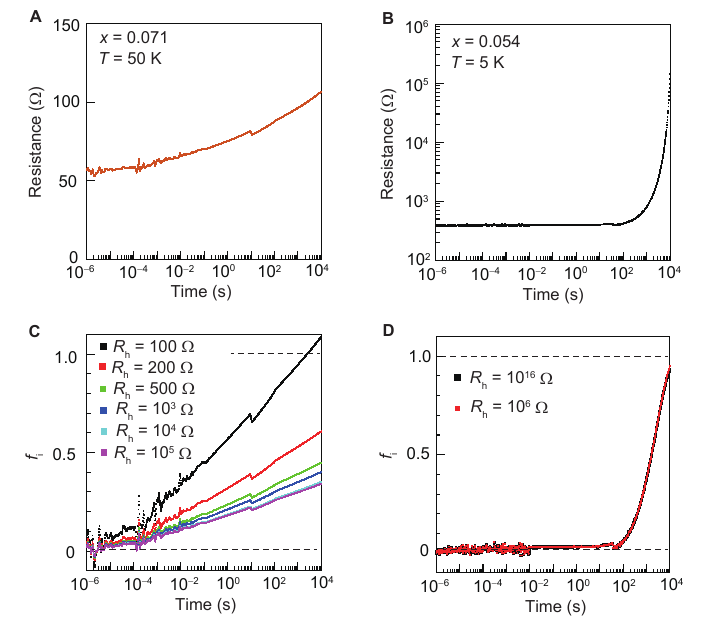} 
		\caption{\textbf{Effect of the uncertainty in $R_h$ on the estimated insulating volume fraction $f_i$.}
			(\textbf{A} and \textbf{B}) Time evolution of the resistance for $x=0.071$ at 50~K (\textbf{A}) and for $x=0.054$ at 5~K (\textbf{B}).
			(\textbf{C} and \textbf{D}) Time evolution of the estimated insulating volume fraction $f_i$ calculated from the data shown in \textbf{A} (\textbf{C}) and \textbf{B} (\textbf{D}) for several values of $R_h$.}
		\label{fig:sup_gem} 
	\end{figure}

For $R_h$, the saturated resistance value was used when the relaxation reached saturation within $10^4$~s. When the relaxation was still ongoing at $10^4$~s, the relaxation was first allowed to complete at a temperature where it proceeded sufficiently fast, and $R_h$ was then determined after subsequently changing the temperature to the target value. For $x=0.071$, the relaxation did not reach saturation at any measured temperature (Fig.~\ref{fig:relaxation}C and Fig.~\ref{fig:sup_gem}A), and thus $R_h$ could not be determined directly from a saturated resistance value. In the relaxation measurements, the resistance directly observed within the accessible time window remained below 150~$\Omega$ but exhibited a continuous increase without saturation. Assuming that the low-temperature equilibrium state is insulating, $R_h$ is expected to be substantially larger than the experimentally observed resistance. Based on this assumption, a representative value of $R_h=1000~\Omega$ was adopted in the analysis. The estimated $f_i$ showed a substantial dependence on $R_h$ between 100~$\Omega$ and 1000~$\Omega$, whereas the dependence became much weaker for $R_h \geq 1000~\Omega$ (Fig.~\ref{fig:sup_gem}C). For $x=0.054$, although the resistance in the low-temperature insulating state exceeded the measurable range ($\sim10^9~\Omega$), the estimated insulating fraction $f_i$ showed only minor deviations when calculated using $R_h=10^6~\Omega$ and $10^{16}~\Omega$, indicating that the estimation is robust against variations in $R_h$ (Fig.~\ref{fig:sup_gem}B,D).

\subsubsection*{Phase-field simulation}
Phase-field simulations were performed to visualize the spatial evolution of the phase transformation. The system was described by a non-conserved phase-field variable $\phi$, where $\phi=0$ and $\phi=1$ represent the metastable metallic phase and the stable insulating phase, respectively. The simulations were carried out on a $2000\times2000$ grid with a mesh size of 100~nm using zero-flux boundary conditions. The initial condition consisted of a circular domain of metastable phase with a radius of 50~$\mu$m embedded in the surrounding stable phase ($\phi=1$). The time evolution of $\phi$ was calculated using an Allen--Cahn-type equation,
\[
\frac{\partial \phi}{\partial t}
=
M(T)
\left[
a^2\nabla^2\phi
-
\frac{\partial f}{\partial\phi}
\right],
\]
where $M(T)=M_0\exp(-\Delta_k/T)$ is the temperature-dependent phase-field mobility, and $a$ is the gradient coefficient. The local free-energy density was written as
\[
f(\phi)=
W\phi^2(1-\phi)^2+
\Delta g(T)(10\phi^3-15\phi^4+6\phi^5),
\]
where $\Delta g(T)=\Delta g_0(1-T/T_c)$ is the free-energy density difference between the two phases, and $W$ is the double-well barrier height. At each time step, nucleation was implemented by sampling the number of nucleation events from a Poisson distribution with a mean of $I(T)V\Delta t$, where $I(T)=A\exp(-\Delta_k/T)$, $A$ is the nucleation prefactor, $V$ is the simulation volume, and $\Delta t$ is the time step. The local phase-field variable at each randomly selected nucleation site was set to $\phi=1$, and the subsequent growth step was calculated using the Allen--Cahn equation.

\subsubsection*{Phonon calculations}
Phonon calculations for the rutile phase of V$_{1-x}$W$_x$O$_2$ were performed within the harmonic approximation using the finite-displacement method as implemented in Phonopy \cite{togo2023first}. Interatomic force constants were obtained from forces calculated in a 2 $\times$ 2 $\times$ 2 supercell containing 48 atoms. The force calculations were carried out using the Vienna Ab initio Simulation Package (VASP) \cite{kresse1996efficient} with the projector augmented-wave method \cite{kresse1999ultrasoft} and the Perdew--Burke--Ernzerhof exchange--correlation functional \cite{perdew1996generalized}. The on-site Coulomb interaction was treated within the Dudarev DFT+$U$ approach \cite{dudarev1998electron}. Following a previous DFT+$U$ study of VO$_2$ \cite{kim2013correlation}, we used an effective Hubbard parameter of $U_{\mathrm{eff}} = 3.4$ eV for V 3$d$ states, while $U_{\mathrm{eff}} = 0$ was used for W and O. A plane-wave cutoff energy of 520 eV and a 4 $\times$ 4 $\times$ 8 k-point mesh were used for the 2 $\times$ 2 $\times$ 2 supercell, corresponding to an 8 $\times$ 8 $\times$ 16 mesh for the primitive rutile cell. W substitution was treated using the virtual crystal approximation. Phonon dispersions and densities of states were calculated from the harmonic interatomic force constants.


\subsection*{Supplementary Text}
\subsubsection*{Sub-micrometer phase mixing in V$_{1-x}$W$_x$O$_2$}

\begin{figure} 
		\centering
		\includegraphics[width=0.9\textwidth]{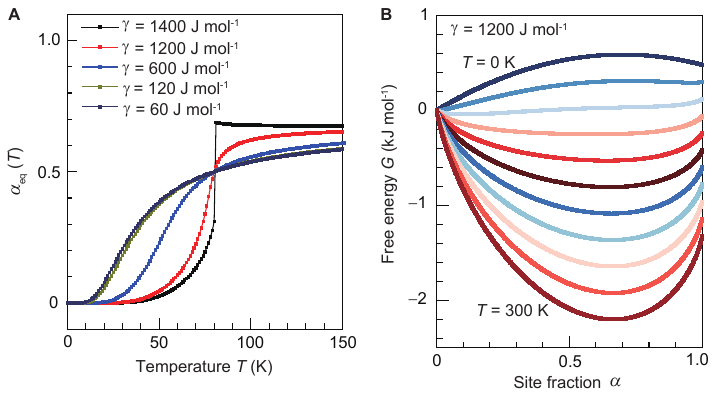} 
		\caption{\textbf{Free-energy analysis based on the Slichter--Drickamer model.}
			(\textbf{A}) Temperature dependence of the equilibrium site fraction $\alpha_{\mathrm{eq}}(T)$ obtained by minimizing the free energy for $\gamma=60$, 120, 600, 1200, and 1400~J~mol$^{-1}$. Larger values of $\gamma$, corresponding to a higher energetic penalty for the coexistence of different local structures, result in a sharper change in $\alpha_{\mathrm{eq}}(T)$. A site fraction of 0 (1) corresponds to all sites having insulating-type (metallic-type) local structures.
			(\textbf{B}) Free-energy landscape $G(\alpha)$ calculated for $\gamma=1200$~J~mol$^{-1}$ at temperatures between 0 and 300~K in steps of 30~K. The equilibrium site fraction $\alpha_{\mathrm{eq}}$ corresponds to the minimum of the free energy at each temperature.}
		\label{fig:sup_sd} 
	\end{figure}

To discuss the energetics of phase mixing, we analyze the temperature-dependent free-energy landscape using the Slichter--Drickamer (SD) model, which is widely employed to describe structural phase transitions \cite{slichter1972pressure,tokoro2015external} (Fig.~\ref{fig:sup_sd}). In this model, the free energy is written as

\begin{equation}
	G=\alpha \Delta H+\gamma \alpha(1-\alpha)+T\left\{R\left[\alpha\ln\alpha+(1-\alpha)\ln(1-\alpha)\right]-\alpha \Delta S\right\},
\end{equation}

where $\alpha$ is the site fraction of the metallic-type local structure, $\Delta H$ and $\Delta S$ are the enthalpy and entropy differences between the two pure local structures, $\gamma$ is the interaction parameter, $T$ is the temperature, and $R$ is the gas constant. The term $R[\alpha \ln \alpha + (1-\alpha)\ln(1-\alpha)]$ represents the configurational entropy associated with the coexistence of the two local structures. In the high-temperature limit, the value of $\alpha$ that minimizes $G$ approaches

\begin{equation}
	\alpha_{\infty}=\frac{1}{1+\exp(-\Delta S/R)}.
\end{equation}

This expression indicates that the asymptotic value of $\alpha$ is governed by the ratio $\Delta S/R$. When $\Delta S \gg R$, the entropy gain associated with the metallic-type local structure dominates and $\alpha_{\infty}$ approaches unity. In contrast, when $\Delta S$ is comparable to or smaller than $R$, the system gains entropy by mixing the two local structures, resulting in an intermediate value of $\alpha_{\infty}$. Previous studies have shown that the entropy change across the metal--insulator transition decreases with W concentration $x$ to approximately 6~J~mol$^{-1}$~K$^{-1}$ at $x=0.05$ \cite{uchimura2020robust}. This value is comparable to $R=8.314$~J~mol$^{-1}$~K$^{-1}$, giving $\alpha_{\infty}\approx0.67$. Therefore, the high-temperature phase of V$_{1-x}$W$_x$O$_2$ is expected to contain a substantial fraction of insulating-type local structures in addition to metallic-type ones. This picture is consistent with X-ray studies on V$_{1-x}$W$_x$O$_2$ films that have revealed weak diffraction intensity associated with the low-temperature crystal structure even above $T_{\mathrm{MI}}$ for $0.05\leq x \leq0.07$ \cite{okuyama2015x}. These results suggest that insulating-type local structures persist as thermal fluctuations above $T_{\mathrm{MI}}$, although static long-range ordering develops only below $T_{\mathrm{MI}}$.

To relate the energetics described by the SD model to the experimentally observed broad metal--insulator transition, we evaluated the temperature dependence of the equilibrium site fraction $\alpha_{\mathrm{eq}}(T)$ by calculating the free-energy landscape within the SD model and minimizing the free energy with respect to $\alpha$. The experimentally determined entropy change may already include contributions associated with local configurational disorder; nevertheless, $\Delta S=6$~J~mol$^{-1}$~K$^{-1}$ was adopted for simplicity. In the present epitaxial films, the enthalpy difference is expected to differ from the bulk value owing to strain effects. We therefore adopted $\Delta H=T_{\mathrm{MI}}\Delta S$, yielding $\Delta H\approx480$~J~mol$^{-1}$. The interaction term $\gamma\alpha(1-\alpha)$ represents the energetic penalty associated with the coexistence of different local structures. Calculations of $\alpha_{\mathrm{eq}}(T)$ for different values of $\gamma$ show that larger $\gamma$ produces a sharper change in $\alpha_{\mathrm{eq}}(T)$ (Fig.~\ref{fig:sup_sd}A). For $\gamma\lesssim1200$~J~mol$^{-1}$, $\alpha_{\mathrm{eq}}$ continues to increase with increasing temperature even above the transition temperature. This behavior is consistent with the experimentally observed temperature dependences of the electrical resistance and lattice parameters above $T_{\mathrm{MI}}$, suggesting that the fraction of metallic-type local structures remains temperature dependent in the high-temperature phase. Thus, within the framework of the SD model, the experimentally observed broad metal--insulator transition reflects a relatively small energetic penalty for the coexistence of distinct local structures.

To examine the origin of the small $\gamma$ inferred from the SD analysis, we discuss the contribution of lattice mismatch to the coexistence energy of different local structures. For V$_{1-x}$W$_x$O$_2$, the elastic modulus is approximately 140~GPa \cite{tsai2004effect}, and the lattice strain associated with the phase transition is approximately 0.5--1\% along the out-of-plane direction \cite{okuyama2015x}. If one of the two coexisting local structures accommodates this lattice mismatch, Hooke's law yields an elastic energy of approximately 50~J~mol$^{-1}$. This energy scale is much smaller than the characteristic free-energy scale of the SD model, which is of the order of 1~kJ~mol$^{-1}$ (Fig.~\ref{fig:sup_sd}B). By comparison, IrTe$_2$, which lies in a crossover regime between nucleation-dominated and growth-dominated transformations \cite{oike2021real}, has a similar elastic modulus ($\sim110$~GPa \cite{li2015synthesis}) but exhibits a much larger lattice strain of approximately 2--3\% \cite{ko2015charge}. Within the same approximation, this corresponds to an elastic energy of the order of 500~J~mol$^{-1}$, an order of magnitude larger than that of V$_{1-x}$W$_x$O$_2$. Such an energy scale is expected to make a substantial contribution to the free-energy landscape. Thus, the small lattice mismatch in V$_{1-x}$W$_x$O$_2$ contributes to the small interaction parameter inferred from the SD analysis.

The SD analysis provides a possible microscopic picture of the metastable state generated by quenching. Because the equilibrium high-temperature phase is predicted to contain a substantial fraction of insulating-type local structures when $\Delta S$ is comparable to $R$, rapid quenching is expected to freeze these local structural motifs into the supercooled state. Such local structures can subsequently act as nuclei for the insulating phase, providing a natural explanation for the nucleation-dominated transformation kinetics observed experimentally. The small interaction parameter $\gamma$ inferred from the SD analysis has an additional implication. Because the energetic penalty associated with the coexistence of different local structures is small, finely intermixed regions can be maintained with only a small free-energy cost. This condition is therefore consistent with the experimentally observed sub-micrometer-scale phase mixing in V$_{1-x}$W$_x$O$_2$.


\begin{figure} 
		\centering
		\includegraphics[width=0.9\textwidth]{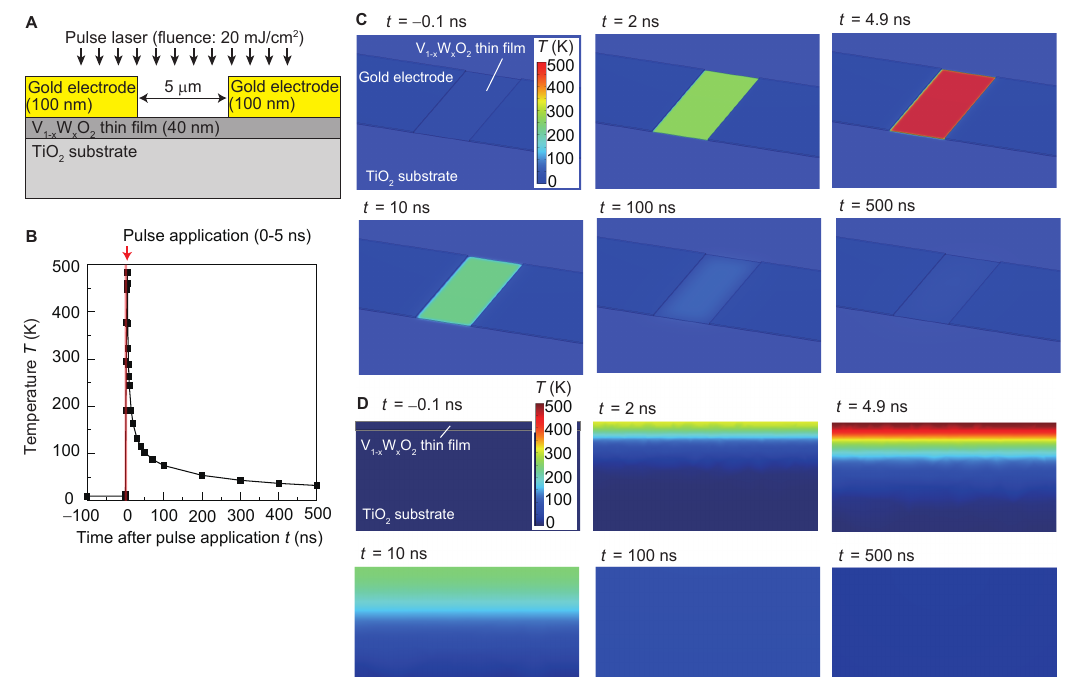} 
		\caption{\textbf{Simulated temperature evolution during pulsed laser irradiation.}
			(\textbf{A}) Schematic illustration of the simulated structure. A 40-nm-thick V$_{1-x}$W$_x$O$_2$ film on a TiO$_2$ substrate with a gold electrode was assumed in the finite-element simulations. Full absorption of the incident laser pulse in the V$_{1-x}$W$_x$O$_2$ layer was assumed.
			(\textbf{B}) Temporal evolution of the temperature at the center of the film. The simulation indicates a temperature rise exceeding $\sim$400 K and a subsequent cooling rate on the order of $10^9$ K s$^{-1}$.
			(\textbf{C} and \textbf{D}) Spatial temperature distributions in the film shown in the top view (\textbf{C}) and cross-sectional view (\textbf{D}). During laser irradiation ($t=0$--5 ns), a steep temperature gradient develops from the thin film toward the substrate. After the laser pulse, the gradient gradually decreases owing to heat flow from the film into the substrate.}
		\label{fig:sup_fem} 
	\end{figure}

\begin{figure} 
		\centering
		\includegraphics[width=0.9\textwidth]{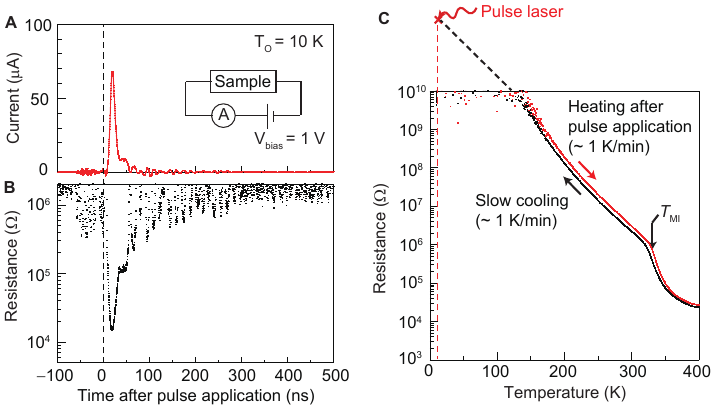} 
		\caption{\textbf{Electrical response of VO$_2$ under pulsed laser irradiation.}
			(\textbf{A} and \textbf{B}) Time dependences of the current (\textbf{A}) and resistance (\textbf{B}) following laser irradiation. The resistance decreases to approximately 10~k$\Omega$ and subsequently recovers to above 1~M$\Omega$ within approximately 100~ns.
			(\textbf{C}) Temperature dependence of the independently measured two-probe resistance. The transient resistance drop corresponds to a temperature increase up to approximately 400~K based on the resistance--temperature relation under slow cooling. The recovery occurs on a $\sim$100~ns timescale, corresponding to the slowest process among the cooling time, the lifetime of the supercooled state, and the instrumental response time, implying that the cooling time is at most $\sim$100~ns. Finite-element simulations (Fig.~\ref{fig:sup_fem}) indicate that the sample temperature exceeds approximately 400~K during the laser pulse and decreases on a comparable timescale. This consistency supports that the experiment operates in an ultrafast thermal-quenching regime.}
		\label{fig:sup_vo2} 
	\end{figure}
	
\begin{figure} 
		\centering
		\includegraphics[width=0.9\textwidth]{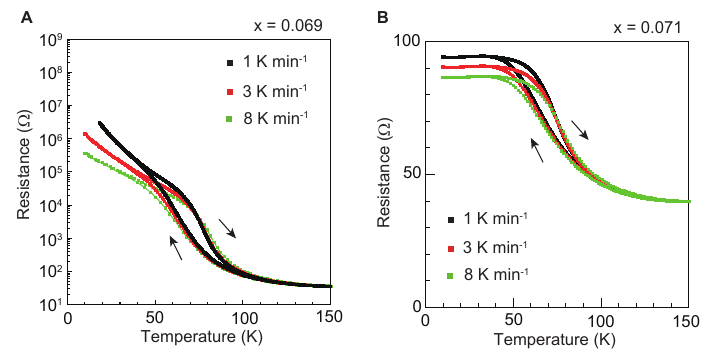} 
		\caption{\textbf{Cooling-rate dependence of the two-probe resistance.}
			(\textbf{A} and \textbf{B}) Temperature dependences of the resistance measured at cooling rates of 1, 3, and 8~K~min$^{-1}$ for $x=0.069$ (\textbf{A}) and $x=0.071$ (\textbf{B}). Below 80~K, the resistance depends on the cooling rate, with faster cooling resulting in lower resistance. This behavior is consistent with the gradual increase in resistance observed under isothermal conditions below approximately 80~K (Fig.~\ref{fig:relaxation}C). These results imply that a metastable metallic state is partially retained under finite cooling rates.}
		\label{fig:sup_ratedep} 
	\end{figure}
	
\begin{figure} 
		\centering
		\includegraphics[width=0.9\textwidth]{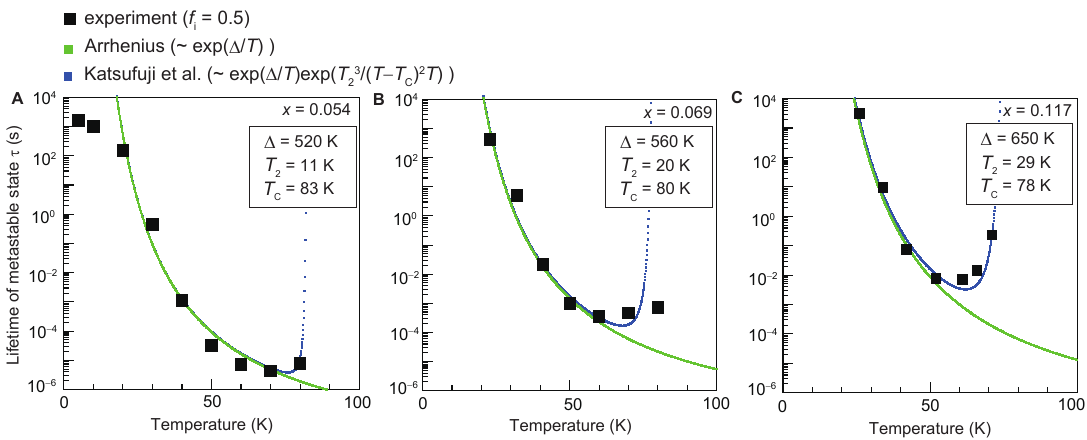} 
		\caption{\textbf{Estimation of the activation barrier for the metastable state.}
			(\textbf{A}--\textbf{C}) Temperature dependences of the time at which the insulating fraction $f_i$ exceeds 0.5 for $x=0.054$ (\textbf{A}), 0.069 (\textbf{B}), and 0.117 (\textbf{C}). The data were fitted using an Arrhenius form and the model proposed by Katsufuji \textit{et al.}~\cite{katsufuji2020nucleation}. While the latter reproduces the non-monotonic temperature dependence, both models yield the same activation barrier in the low-temperature regime, where the relaxation exhibits Arrhenius-like behavior. From these fittings, the activation barrier $\Delta$ is estimated to be in the range of 500--700~K.}
		\label{fig:sup_ttt} 
	\end{figure}
	
\begin{figure} 
		\centering
		\includegraphics[width=0.9\textwidth]{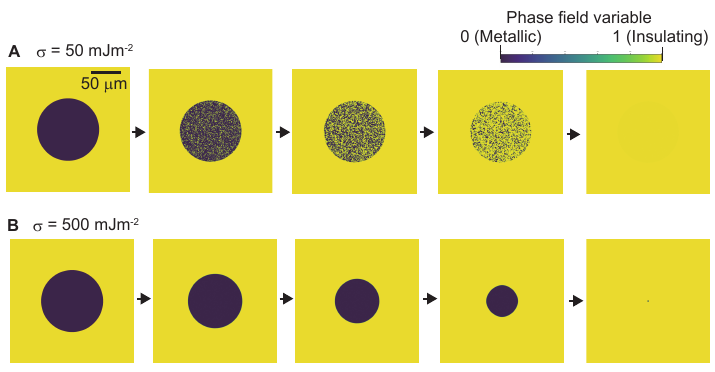} 
		\caption{\textbf{Phase-field simulations of domain evolution.}
			(\textbf{A} and \textbf{B}) Time evolution of the real-space configurations of the phase-field variable $\phi(x,y)$, starting from a circular metastable domain ($\phi=0$) with a radius of 50~$\mu$m embedded in a stable matrix ($\phi=1$), for interfacial energies of 50~mJ~m$^{-2}$ (\textbf{A}) and 500~mJ~m$^{-2}$ (\textbf{B}). Lower interfacial energy leads to nucleation-dominated kinetics, whereas higher interfacial energy leads to growth-dominated kinetics. The former reproduces the experimentally observed spatially homogeneous relaxation on the micrometer scale. The simulations were performed at $T=30$~K. The transition temperature and kinetic barrier were set to $T_c=80$~K and $\Delta_k=600$~K, respectively, close to the corresponding experimental values. The remaining phenomenological parameters were chosen to reproduce the experimental relaxation behavior: $\Delta g_0=-1.0\times10^7$~J~m$^{-3}$, $M_0=10$~m$^3$~J$^{-1}$~s$^{-1}$, $A=1.0\times10^{26}$~s$^{-1}$~m$^{-3}$, $\delta=4.0\times10^{-7}$~m, and $\lambda=0.1$. The gradient coefficient and double-well barrier height were given by $a=(3\delta\sigma/b)^{1/2}$ and $W=6\sigma b/\delta$, respectively, where $b=2\tanh^{-1}(1-2\lambda)$ and $\sigma$ is the interfacial energy. The implications of the nucleation-dominated regime for sub-micrometer phase mixing are discussed in Supplementary Text.}
		\label{fig:sup_pf} 
	\end{figure}
	
\begin{figure} 
		\centering
		\includegraphics[width=0.9\textwidth]{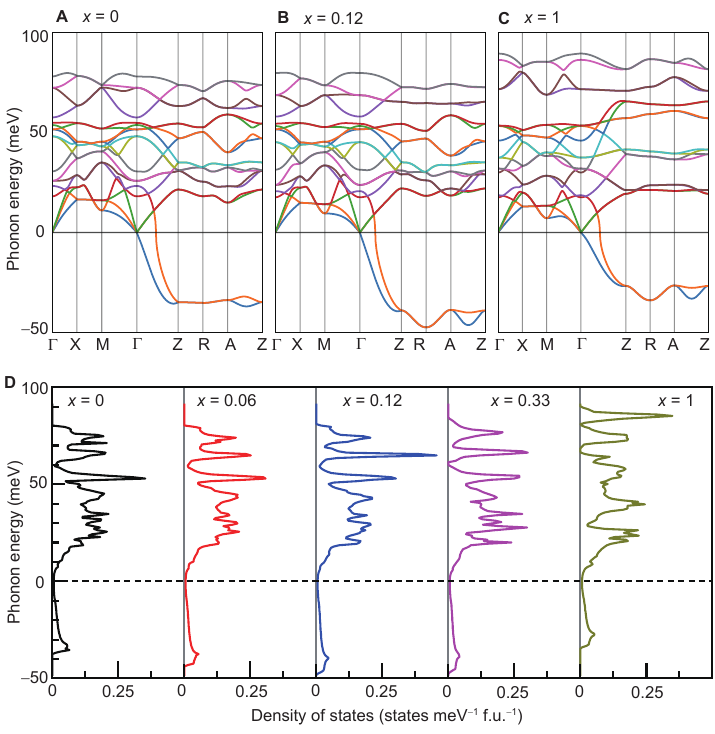} 
		\caption{\textbf{Phonon calculations for rutile V$_{1-x}$W$_x$O$_2$.}
			(\textbf{A}--\textbf{C}) Phonon dispersions calculated for $x=0$ (\textbf{A}), $x=0.12$ (\textbf{B}), and $x=1$ (\textbf{C}). The high-symmetry points are defined in reciprocal lattice units as $\Gamma=(0,0,0)$, X=$(1/2,0,0)$, M=$(1/2,1/2,0)$, Z=$(0,0,1/2)$, R=$(1/2,0,1/2)$, and A=$(1/2,1/2,1/2)$. (\textbf{D}) Phonon density of states for $0\leq x\leq1$. The intermediate compositions were treated within the virtual crystal approximation. The DFT+$U$ calculations were performed with $U_{\mathrm{eff}}=3.4$~eV for V 3$d$ states, while $U_{\mathrm{eff}}=0$ was used for W and O. For $x=0$ (VO$_2$), imaginary phonon modes appear along the Z--R--A--Z path, reproducing previous first-principles results~\cite{kim2013correlation}. For $x=1$ (WO$_2$), imaginary modes also appear along the Z--R--A--Z path with $U_{\mathrm{eff}}=0$. For intermediate compositions treated within the virtual crystal approximation, these instabilities persist, indicating that the rutile structure remains dynamically unstable within the harmonic approximation.}
		\label{fig:sup_phonon} 
	\end{figure}
	
\begin{table}
	\centering
	\caption{\textbf{Resistance values used in the GEM analysis.}
		Values of $R_l$ and $R_h$ used to estimate the insulating volume fraction
		for each composition and temperature.}
	\label{tab:GEM_parameters}
	
	\begin{tabular}{ccc@{\hspace{1.0cm}}ccc}
		\hline
		\multicolumn{3}{c}{$x=0.054$}
		&
		\multicolumn{3}{c}{$x=0.069$}\\
		\hline
		$T$ (K) & $R_h$ ($\Omega$) & $R_l$ ($\Omega$)
		&
		$T$ (K) & $R_h$ ($\Omega$) & $R_l$ ($\Omega$)\\
		\hline
		5  & $1.1\times10^{16}$ & 380
		&
		15 & $6.3\times10^{7}$ & 65\\
		10 & $4.2\times10^{12}$ & 380
		&
		25 & $6.3\times10^{7}$ & 65\\
		20 & $5.7\times10^{9}$ & 380
		&
		32 & $3.0\times10^{6}$ & 65\\
		30 & $2.0\times10^{8}$ & 380
		&
		41 & $3.5\times10^{6}$ & 65\\
		40 & $2.2\times10^{7}$ & 380
		&
		50 & $2.4\times10^{5}$ & 65\\
		50 & $4.5\times10^{6}$ & 380
		&
		60 & $6.7\times10^{4}$ & 65\\
		60 & $1.3\times10^{6}$ & 380
		&
		70 & $1.5\times10^{4}$ & 65\\
		70 & $2.5\times10^{5}$ & 380
		&
		80 & $1.1\times10^{3}$ & 65\\
		80 & $4.6\times10^{4}$ & 380
		&
		&                       &   \\
		\hline
		\\[-0.6em]
		\multicolumn{3}{c}{$x=0.071$}
		&
		\multicolumn{3}{c}{$x=0.117$}\\
		\hline
		$T$ (K) & $R_h$ ($\Omega$) & $R_l$ ($\Omega$)
		&
		$T$ (K) & $R_h$ ($\Omega$) & $R_l$ ($\Omega$)\\
		\hline
		30 & $1.0\times10^{3}$ & 55
		&
		21 & $1.7\times10^{9}$ & 56\\
		40 & $1.0\times10^{3}$ & 55
		&
		26 & $2.0\times10^{8}$ & 56\\
		50 & $1.0\times10^{3}$ & 55
		&
		34 & $3.1\times10^{7}$ & 56\\
		60 & $1.0\times10^{3}$ & 55
		&
		42 & $6.8\times10^{6}$ & 56\\
		70 & $1.0\times10^{3}$ & 55
		&
		52 & $3.5\times10^{6}$ & 56\\
		&                       &
		&
		61 & $4.0\times10^{5}$ & 56\\
		&                       &
		&
		66 & $1.9\times10^{5}$ & 56\\
		&                       &
		&
		71 & $7.6\times10^{4}$ & 56\\
		\hline
	\end{tabular}
\end{table}




\end{document}